\documentclass[useAMS,usenatbib,galley]{mn2e}
\pdfoutput=1
\usepackage{bm}
\usepackage{natbib}
\usepackage{graphicx}
\usepackage{color}
\usepackage{amsmath}
\usepackage{amssymb}

\newcommand{\aap}{Astron.\ Astrophys.\ }
\newcommand{\apj}{Astrophys.\ J.\ }
\newcommand{\apjl}{Astrophys.\ J.\ Lett.\  }
\newcommand{\apjs}{Astrophys.\ J.\ Suppl.\  }
\newcommand{\mnras}{Mon.\ Not.\ R.\ Astron.\ Soc.\ }
 
\newcommand{\physrep}{Phys.\ Rep.\  }

\newcommand{\prd}{Phys.\ Rev.\ D\ }

\newcommand{\ie}{{\it i.e.~}}

\newcommand{\eqn}[1]{equation~(\ref{#1})}
\newcommand{\Eqn}[1]{Equation~(\ref{#1})}
\newcommand{\fig}[1]{Fig.~\ref{#1}}

\newcommand{\beq}{\begin{equation}}
\newcommand{\eeq}{\end{equation}}
\newcommand{\bdm}{\begin{displaymath}}
\newcommand{\edm}{\end{displaymath}}
\newcommand{\bea}{\begin{eqnarray}}
\newcommand{\eea}{\end{eqnarray}}
\newcommand{\bt}{\begin{tabular}}
\newcommand{\et}{\end{tabular}}
\newcommand{\xv}{{\bf x}}
\newcommand{\kv}{{\bf k}}

\newcommand{\qv}{{\bf q}}

\newcommand{\intq}{\int\!\!d^3 q}

\def\d{\delta}
\def\D{\Delta}

\def\Mpc{\, h^{-1} \, {\rm Mpc}}

\def\cGpc{\, h^{-3} \, {\rm Gpc}^3}
\def\kMpc{\, h \, {\rm Mpc}^{-1}}

\def\fNL{f_{NL}}

\def\la{\langle}
\def\ra{\rangle}
\def\O{\mathcal O}
\def\kall{k_1,k_2,k_3}

\title[The matter bispectrum and non-Gaussian initial conditions]{The matter bispectrum in N-body simulations with non-Gaussian initial conditions}
\author[E. Sefusatti, M. Crocce and V. Desjacques]{E. Sefusatti$^{1}$\thanks{E-mail:
emiliano.sefusatti@cea.fr (ES)}, M. Crocce$^{2}$\thanks{E-mail: crocce@ieec.uab.es (MC)} and V. Desjacques$^{3}$\thanks{E-mail: dvince@physik.uzh.ch (VD)}\\
$^{1}$Institut de Physique Th\'eorique, CEA, IPhT, F-91191 Gif-sur-Yvette, France\\
$^{2}$Institut de Ci\`encies de l'Espai, IEEC-CSIC, Campus UAB, Facultat de Ci\`encies, Torre C5 par-2, Barcelona 08193, Spain\\
$^{3}$Institute for Theoretical Physics, Universit\"at Z\"urich, Winterthurerstrasse 190, CH-8057 Z\"urich, Switzerland}

\begin{document}

%
%
%


\pagerange{\pageref{firstpage}--\pageref{lastpage}} 

\maketitle
\label{firstpage}

\begin{abstract}
We present measurements of the dark matter bispectrum in N-body simulations with non-Gaussian initial conditions of the local kind for a large variety of triangular configurations and compare them with predictions from Eulerian Perturbation Theory up to one-loop corrections. We find that the effects of primordial non-Gaussianity at large scales, when compared to Perturbation Theory,  are well described by the initial component of the matter bispectrum, linearly extrapolated at the redshift of interest. In addition, we find that, for $\fNL=100$, the nonlinear corrections {\em due} to non-Gaussian initial conditions are of the order of $\sim 3$-$4$\% for generic triangles up to $\sim 20$\% for squeezed configurations, {\em at any redshift}. We show that the predictions of Perturbation Theory at tree-level fail to describe the simulation results at redshift $z=0$ at scales corresponding to $k\sim 0.02$-$0.08\kMpc$, depending on the triangle, while one-loop corrections can significantly extend their validity to smaller scales. At higher redshift, one-loop Perturbation Theory provides indeed quite accurate predictions, particularly with respect to the {\em relative} correction due to primordial non-Gaussianity.
\end{abstract}




\section{Introduction}

In recent years, a significant research activity has been devoted to the effects of a possible small departure from Gaussianity in the primordial cosmological perturbations. While current constraints on primordial non-Gaussianity from measurements of the cosmic microwave background (CMB) and the large-scale structure are still consistent with the Gaussian hypothesis \citep{KomatsuEtal2010, SlosarEtal2008}, a possible detection in forthcoming experiments would constitute a major discovery, providing crucial information on the early Universe and on the high-energy physics of inflation \citep[see, for instance,][]{KomatsuEtal2009A}. 

The effect of primordial non-Gaussianity on the large-scale structure has been assumed, for a long time, to be limited to an additional, {\em primordial} component to the matter skewness and bispectrum induced by gravitational instability and to a correction to the abundance of massive cluster \citep[see][for recent reviews]{LiguoriEtal2010, DesjacquesSeljak2010B}. Numerical and analytical studies have indeed shown that a matter density probability distribution initially skewed toward positive values produces more overdense regions and, consequently, collapsed objects while a negatively skewed distribution produces larger voids \citep[see][for recent work]{GrossiEtal2008, PillepichPorcianiHahn2010, KamionkowskiVerdeJimenez2009, LamSheth2009, LamShethDesjacques2009, MaggioreRiotto2009C}. Moreover, a nonvanishing skewness in the initial conditions corresponds to a primordial component to the matter bispectrum, \ie the three-point function in Fourier space. For the {\em local} non-Gaussian model considered here, the primordial matter bispectrum exhibits a scale, redshift and triangle shape dependence distinct from that of the component sourced by the nonlinear growth of structures. This enables us in principle to disentangle the two contributions. In the specific case of equilateral triangular configurations, the primordial contribution to the matter bispectrum scales as $\sim k^{-2}$ relative to the gravity-induced term, leading to large, potentially observable corrections at low wavenumbers. Measurements of the galaxy bispectrum in future large-volume redshift surveys (such as Euclid or HETDEX) should be able to provide constraints on the local non-Gaussian model competitive with those from CMB observations \citep{ScoccimarroSefusattiZaldarriaga2004, SefusattiKomatsu2007, SefusattiEtal2009}. 

In addition to these effects, \citet{DalalEtal2008} have recently discovered a large correction to the galaxy bias in numerical simulations of {\em local} primordial non-Gaussianity. Further numerical and theoretical work has confirmed this result \citep{MatarreseVerde2008, SlosarEtal2008, AfshordiTolley2008, McDonald2008, GrossiEtal2008, TaruyaKoyamaMatsubara2008, PillepichPorcianiHahn2010, DesjacquesSeljakIliev2009, GiannantonioPorciani2009}. The constraints obtained from power spectrum measurement of highly biased objects in current data-sets are already comparable to the CMB results \citep{SlosarEtal2008, DesjacquesSeljak2010}, and the prospects for detecting local primordial non-Gaussianity with galaxy clustering look exciting \citep{DalalEtal2008, CarboneVerdeMatarrese2008, Seljak2009, Slosar2009, VerdeMatarrese2009, DesjacquesSeljak2010}. At this point, analyses of the galaxy bispectrum preceeding the work of \citet{DalalEtal2008} must be updated to account for the non-Gaussian correction to the galaxy bias. In fact, a rigorous {\em joint} analysis of the galaxy power spectrum and bispectrum in presence of local non-Gaussianity is in order. First steps in this direction have been taken by \citet{JeongKomatsu2009B} and \citet{Sefusatti2009} with a preliminary comparison with simulations in \citet{NishimichiEtal2009}. 

In this perspective, we will consider the measurement of several triangular configurations of the {\em matter} bispectrum on mildly nonlinear scales, with both Gaussian and non-Gaussian initial conditions of the local type. Although the matter bispectrum is not directly observable with tracers of the large-scale structure, it is instructive to assess the extent to which perturbation theory describes the shape dependence of the matter three-point function in the presence of non-Gaussianity of the local type. This analysis will be useful when considering the complication brought by biasing, which will be addressed in a forthcoming publication. Measurements of the matter power spectrum with local non-Gaussianity can be found in \citet{PillepichPorcianiHahn2010, DesjacquesSeljakIliev2009}, where the small corrections at mildly nonlinear scales predicted in the framework of perturbation theory by \citet{TaruyaKoyamaMatsubara2008} are observed. In the case of the matter bispectrum, measurements in simulations with Gaussian initial conditions are shown in \citet{ScoccimarroEtal1998, HouEtal2005, PanColesSzapudi2007, SmithShethScoccimarro2008, GuoJing2009A}, with \citet{SmithShethScoccimarro2008} considering, in addition, redshift space predictions in the context of the halo model. By contrast, the only measurement so far of the matter (and halo) bispectrum in simulations with local non-Gaussian initial conditions can be found in \citet{NishimichiEtal2009}, where a relatively small subset of isosceles triangular configurations is considered. 

We will compare our measurements with predictions of the matter bispectrum at the one-loop approximation in Eulerian perturbation theory. A comparison of one-loop results with the bispectrum extracted from simulations with Gaussian initial conditions is shown in \citet{ScoccimarroEtal1998}, whereas a comparison of the effect of primordial non-Gaussianity with the tree-level prediction of perturbation theory is performed in \citet{NishimichiEtal2009} for ``squeezed'' isosceles configurations at $z=0$ with $k\lesssim 0.1\kMpc$ only. Here, we will extend the analysis to include several triangular configurations covering the range of scales $0.002\lesssim k \lesssim 0.3\kMpc$ and redshifts $z=0$, $1$ and $2$. This will allow us to broadly test the accuracy of one-loop perturbation theory in the mildly nonlinear regime. We will also discuss the validity of two phenomenological prescriptions for the nonlinear bispectrum with Gaussian initial conditions, namely the fitting function of \citet{ScoccimarroCouchman2001} and the formula of \citet{PanColesSzapudi2007} based on a scaling transformation.

This paper is organized as follows. In section~\ref{sec:theory} we summarize previous results on the predictions of the matter power spectrum and bispectrum in cosmological perturbation theory for both Gaussian and local non-Gaussian initial perturbations. In section~\ref{sec:sims} we describe the N-body simulations and the bispectrum estimator employed in our analysis whereas, in section~\ref{sec:results}, we present our measurements of the matter bispectrum and compare them to one-loop predictions in perturbation theory. Finally, we conclude in section~\ref{sec:conclusions}.

\section{Theory}
\label{sec:theory}

In this section, we summarize previous results on the nonlinear evolution of the matter correlators as described specifically by Eulerian Perturbation Theory (PT). The quantity of interest, the matter overdensity $\d$, is obtained as a perturbative solution to the continuity and Euler equations, and Poisson equation relating matter perturbations and the gravitational potential. These equations fully determine the evolution of the matter density and velocity fields, once the initial conditions are given in terms of the primordial correlators. Other approaches such as Lagrangian Perturbation Theory, for instance, have also been studied in the literature. We refer the reader to \citet{Scoccimarro2000B} for a study of the matter bispectrum in Lagrangian Perturbation Theory with Gaussian initial conditions and to \citet{BernardeauEtal2002} for a complete review of cosmological perturbation theory.

\subsection{Initial conditions}
\label{sec:ICs}

Our N-body simulations of the matter density evolution assume {\em local} non-Gaussian initial conditions. This model of primordial non-Gaussianity is defined by the {\em local} expression in position space for the Bardeen's curvature perturbations $\Phi$ \citep{SalopekBond1990, SalopekBond1991, GanguiEtal1994, VerdeEtal2000, KomatsuSpergel2001}
\beq
\label{eq:phiNG}
\Phi(\xv) = \phi(\xv) + \fNL \left[ \phi^2(\xv) - \langle \phi^2(\xv) \rangle\right],
\eeq
where the second term on the r.h.s. represents a non-Gaussian correction to the Gaussian random field $\phi(\xv)$. In this expression, we assume that $\Phi(\xv)$ is the curvature field during early matter domination, and not the linearly extrapolated value at present time. Despite its relatively simple form, the parameterization of primordial non-Gaussianity provided by \eqn{eq:phiNG} well describes inflationary models in which the non-Gaussianity is produced by local mechanisms on superhorizon scales \citep[see][and references therein]{BartoloEtal2004, LiguoriEtal2010, Chen2010, ByrnesChoi2010}.

The definition of \eqn{eq:phiNG} corresponds to a very specific functional form of the bispectrum and trispectrum of the initial curvature perturbations. One finds the following  leading contribution to the curvature bispectrum,
\beq
\label{eq:Blc}
B_\Phi(k_1, k_2, k_3) = 2 \fNL P_\Phi(k_1)P_\Phi(k_2)+ {\rm 2~perm.}\,,
\eeq  
with the curvature power spectrum $P_\Phi(k)$ defined in terms of the Gaussian component alone as $\langle\phi({\bf k}_1) \phi({\bf k}_2) \rangle = \delta_D^{(3)}(\kv_{12}) P_\Phi(k_1)$, where we introduce the notation $\kv_{ij}\equiv\kv_i+\kv_j$. The magnitude of the curvature bispectrum is maximized for ``squeezed'' triangular configuration, \ie when one side of the triangle is much smaller than the other two, say $k_1\ll k_2\simeq k_3$. The curvature trispectrum is given by,
\bea
\label{eq:Tlc}
T_\Phi(\kv_1,\kv_2,\kv_3,\kv_4) & = & 4 \fNL^2 P_\Phi(k_1)P_\Phi(k_2)\times\nonumber\\
&   &  \left[P_\Phi(k_{13})+P_\Phi(k_{14})\right]+ {\rm 5~perm.}
\eea

The linear matter overdensity in Fourier space $\d_\kv$ is related to the curvature perturbations $\Phi_\kv$ via the Poisson equation,
\beq
\d_\kv(z)=M(k,z)~\Phi_\kv\,,
\eeq
where we introduced the function
\beq
M(k,z)=\frac{2}{3}\frac{k^2T(k)D(z)}{\Omega_mH_0^2}\,,
\eeq
with $T(k)$ being the matter transfer function, computed with the \texttt{CAMB} code \citep{LewisChallinorLasenby2000}, and $D(z)$ the growth factor in units of $1+z$. The initial matter correlators are related to the correlators of the curvature perturbations through
\beq
\la\d_{\kv_1}\cdots\d_{\kv_n}\ra=M(k_1,z)\cdots M(k_n,z)\la\Phi_{\kv_1}\cdots\Phi_{\kv_n}\ra\,,
\eeq
so that the linear power spectrum is given by 
\beq
P_0(k)=M^2(k,z)P_\Phi(k)\,,
\eeq
while the initial bispectrum and trispectrum are given respectively by
\beq\label{eq:B0}
B_0(k_1,k_2,k_3)=M(k_1)M(k_2)M(k_3)B_\Phi(k_1,k_2,k_3)\,,
\eeq
and
\bea\label{eq:T0}
T_0(\kv_1,\kv_2,\kv_3,\kv_4) &\!\!\!\! =\!\!\!\! & M(k_1)M(k_2)M(k_3)M(k_4)\times \nonumber\\
& & T_\Phi(\kv_1,\kv_2,\kv_3,\kv_4)\,.
\eea
As we will see shortly, nonlinear corrections to the matter bispectrum will depend on both the initial bispectrum $B_0$ {\em and} trispectrum $T_0$. 

\subsection{Perturbation theory}
\label{sec:PT}

In PT, the solution for the nonlinear matter density contrast $\d_\kv$ in Fourier space is given in terms of corrections to the linear solution $\d^{(1)}$ \citep{Fry1984}, so that
\beq
\label{eq:PT}
\d_\kv=\d_\kv^{(1)}+\d_\kv^{(2)}+\d_\kv^{(3)}+\ldots.\,,
\eeq
where each nonlinear correction given by
\beq
\label{eq:PTterm}
\d_\kv^{(n)}\equiv\int d^3q_1\ldots d^3q_nF_n(\qv_1,\ldots,\qv_n)~\d_{\qv_1}^{(1)}\ldots\d_{\qv_n}^{(1)}\,,
\eeq
with $F_n(\qv_1,\ldots,\qv_n)$ the {\em symmetrized} kernel of the $n$-order solution. \Eqn{eq:PT} allows one to derive the evolved matter correlators once the initial correlators, \ie the correlators of the linear $\d_\kv^{(1)}$, are known. For Gaussian initial conditions, only the linear power spectrum $P_0$ must be specified. In general however, higher-order correlators need to be taken into account.

In analogy with quantum field theory, perturbative solutions for the matter correlators can be denoted as {\em tree-level} or {\em one-loop}, {\em two-loop}, etc., according to the number of internal integrations present in their expressions. However, it should be noted that, while in the case of Gaussian initial conditions the number of loops of the perturbative correction correspond univocally to a specific perturbative order, this is, as we will see below, no longer true for non-Gaussian initial conditions.  
 
For completeness, we summarize here the explicit expressions of the one-loop PT expansion for both the matter power spectrum and bispectrum with generic non-Gaussian initial conditions. In the case of the matter power spectrum, we have up to fourth order in PT \citep[see][and reference therein]{BernardeauEtal2002}
\bea
\label{eq:Pexp}
P(k) & = & P_{11}(k)+P_{12}(k)+P_{22}^I+P_{13}^I+\nonumber\\
& & {\rm two\!\!-\!\!loop~terms}+\O({\d_0^5}),
\eea 
where, $P_{11}\equiv P_0$ is the linear matter power spectrum, while 
\bea
P_{12}
& = &
2\intq F_2(\qv,\kv-\qv)~B_0(k,q,|\kv-\qv|),
\\
P_{22}^I
& = & 
2\intq F_2^2(\qv,\kv-\qv)~P_0(q)~P_0(|\kv-\qv|),
\\
P_{13}^I
& = & 
6~P_0(k)\intq F_3(\kv,\qv,-\qv)~P_0(q).
\eea
We can see that the only additional contribution due to primordial non-Gaussianity is $P_{12}(k)$ which depends on the initial bispectrum $B_0$ (neglecting two-loop contributions at the fourth order in PT that depend on the initial trispectrum). The amplitude of this correction for local non-Gaussian initial conditions was studied in \citet{TaruyaKoyamaMatsubara2008}, who considered also initial conditions of the equilateral kind.  They found that the effect of a primordial non-Gaussian component within the bounds from CMB observations is typically below 1\% at mildly nonlinear scales, at the limit of detectability in future large-scale structure observations. 

One-loop corrections to the matter bispectrum for Gaussian initial conditions have been studied in \citet{Scoccimarro1997, ScoccimarroEtal1998} while the extension to generic non-Gaussian initial conditions is explored in \citet{Sefusatti2009}. For the bispectrum up to sixth-order in PT and excluding two-loop corrections, we have the following expression
\bea
\label{eq:Bexp}
B & = &  B_{111}+
B_{112}^I+B_{112}^{II}+\nonumber\\
& & 
B_{122}^I+B_{122}^{II}+B_{113}^I+B_{113}^{II}+\nonumber\\
& & 
B_{222}^I+B_{123}^{I}+B_{123}^{II}+B_{114}^{I}+
{\rm 2\!\!-\!\!loop~terms},
\eea
where $B_{111}\equiv B_0$ is the initial bispectrum and 
\beq\label{eq:B112I}
B_{112}^I  =  
2~F_2(\kv_1,\kv_2)~P_0(k_1)~P_0(k_2)+{\rm 2~perm.},
\eeq
is the other tree-level contribution, while the 1-loop corrections are given by
\bea
B_{112}^{II}
& = & 
\intq~ F_2(\qv,\kv_3\!-\!\qv)~T_0(\kv_1,\kv_2,\qv,\kv_3\!-\!\qv),
\label{eq:B112II}\\
B_{122}^{I}
&=&
2 ~P_0(k_1)\left[F_2(\kv_1,\kv_3)\intq~F_2(\qv,\kv_3\!-\!\qv)~
\right.
\times \nonumber\\ & & 
\left.
B_0(k_3,q,|\kv_3-\qv|)+(k_3\leftrightarrow k_2)\right]+{\rm 2~perm.}
\nonumber\\
& = &
F_2(\kv_1,\kv_2)\left[P_0(k_1)~P_{12}(k_2)+P_0(k_2)~P_{12}(k_1)\right]+
\nonumber\\ & & 
{\rm 2~perm.},
\\
B_{122}^{II}
&=&
4 \intq~F_2(\qv,\kv_2\!-\!\qv)~F_2(\kv_1\!+\!\qv,\kv_2\!-\!\qv)~
\times \nonumber\\ & & 
B_0(k_1,q,|\kv_1\!+\!\qv|)~P_0(|\kv_2\!-\!\qv|)
 \nonumber\\ & & 
 +{\rm 2~perm.},
\\
B_{113}^I
&=&
3B_0(k_1,k_2,k_3)\intq~F_3(\kv_3,\qv,-\qv)P_0(q)+
 \nonumber\\ & & 
{\rm 2~perm.},
\\
B_{113}^{II}
&\!=\!&
3 P_0(k_1)\!\!\intq~F_3(\kv_1,\qv,\kv_2\!-\!\qv)B_0(k_2,q,|\kv_2\!-\!\qv|)+
 \nonumber\\ & & 
(k_1\leftrightarrow k_2)+{\rm 2~perm.},
\\
B_{222}^I
\!&\!=\!&\!
8 \!\!\intq F_2(-\qv,\qv\!+\!\kv_1)F_2(-\!\qv\!-\!\kv_1,\qv\!-\!\kv_2)
\times\nonumber\\ & & \!
F_2(\kv_2\!-\!\qv,\qv)P_0(q)P_0(|\kv_1\!+\!\qv|)P_0(|\kv_2\!-\!\qv|),
\\
B_{123}^{I}
\!&\!=\!&\!
6~P_0(k_1)\!\! \intq ~F_3(\kv_1,\kv_2\!-\!\qv,\qv)~F_2(\kv_2\!-\!\qv,\qv)~
\times\nonumber\\ & & 
\!P_0(|\kv_2\!-\!\qv|)~P_0(q)+{\rm 5~perm.},
\\
B_{123}^{II}
\!&\!=\!&\!
6~P_0(k_1)~P_0(k_2)~F_2(\kv_1,\kv_2)
\times\nonumber\\ & & \!
\int d^3q~ F_3(\kv_1, \qv,-\!\qv)~P_0(q)+{\rm 5~perm.}
\nonumber\\
&=&
F_2(\kv_1,\kv_2)\left[P_0(k_1)~P_{13}(k_2)+P_0(k_2)~P_{13}(k_1)\right]+
\nonumber\\ & & 
{\rm 2~perm.},
\\
B_{114}^I
&\!=\!&\!
12\,P_0(k_1)\,P_0(k_2)\!\! \intq \,F_4(\!\qv,\!-\qv,\!-\kv_1,\!-\kv_2)\,P_0(q)+
\nonumber\\ & & {\rm 2~perm.}.
\eea 
Specifically, the one-loop contributions present because of non-Gaussian initial conditions are $B_{112}^{II}$, which depends on the initial trispectrum $T_0$, and all the fifth-order terms $B_{122}^I$, $B_{122}^{II}$, $B_{113}^I$ and $B_{113}^{II}$, which depend on the initial bispectrum $B_0$. 

The remaining terms, corresponding to Gaussian initial conditions, were recently studied in the context of resummation techniques of the PT series and can be regarded as perturbative expansions of ``resummed'' kernels \citep{BernardeauCrocceScoccimarro2008}. For instance $B_{123}^{II}$ corresponds to the next-to-leading term in the resummation of the nonlinear propagator in language of \citet{CrocceScoccimarro2006A, CrocceScoccimarro2006B} or $\Gamma^{(1)}$ in the notation of \citet{BernardeauCrocceScoccimarro2008}. That is, $B_{123}^{II}$ can be obtained from the tree-level expression in \eqn{eq:B112I} by replacing the linear growth $D^2_+$ implicit in $P_0$ as $D^2_+ \rightarrow D^2_+ (1+ P_{13}/P_0)$. Similarly, $B_{114}^I$ corresponds to a redefinition (or re-summation) of the $F_2$ kernel. In turn, $B_{123}^I$ and $B_{222}^I$ are leading terms whose corrections appear at higher order in the PT series of \eqn{eq:Bexp}. The resummed kernels have well defined properties in terms of tree-level quantities and might be the window to an accurate description of the non-linear bispectrum at nonlinear scales. 

\begin{figure*}
{\includegraphics[width=0.95\textwidth]{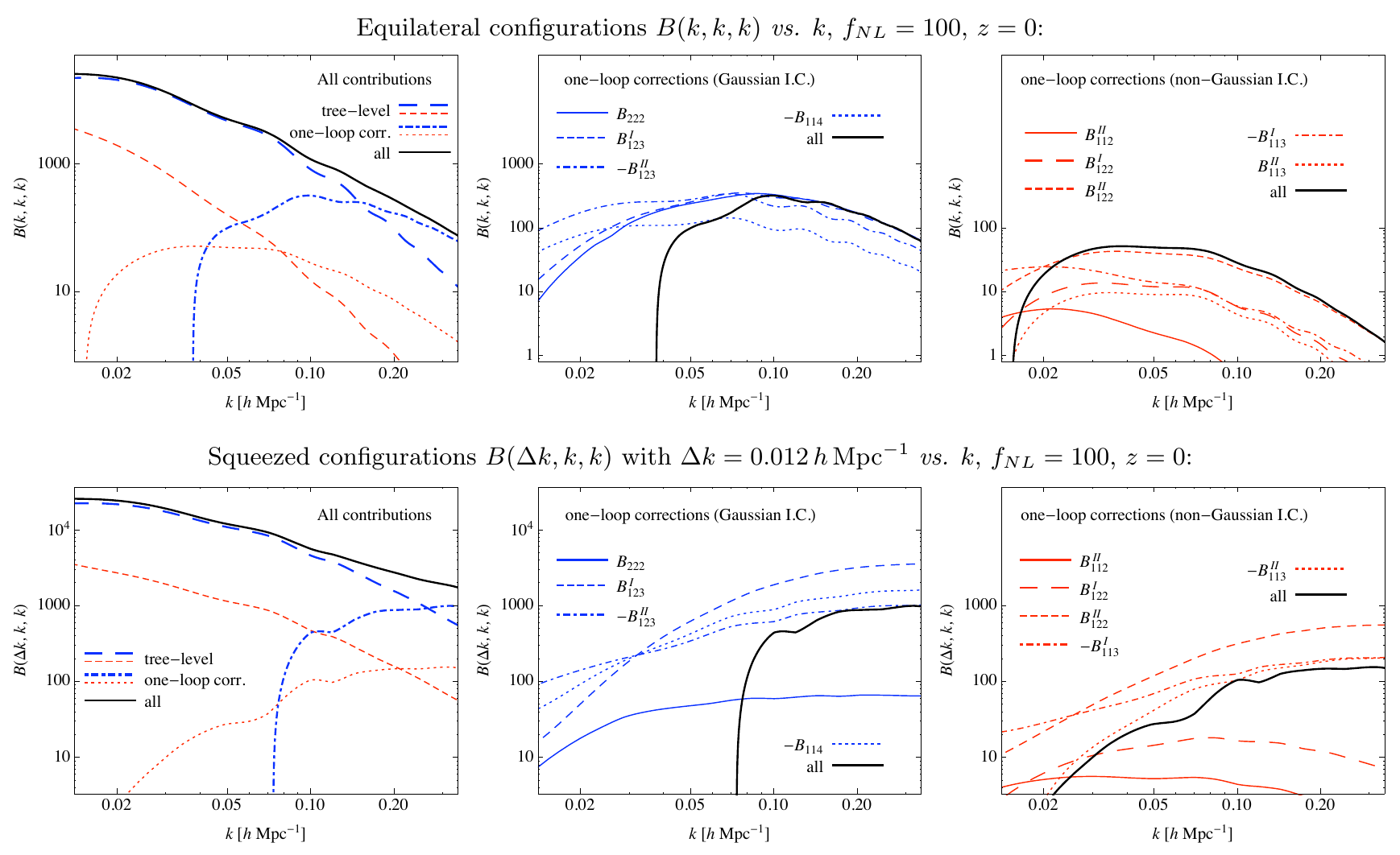}}
\caption{Different components to the PT one-loop prediction for the equilateral configurations $B(k,k,k)$ ({\em upper panel}) and the squeezed configurations $B(\D k,k,k)$ with $\D k\simeq 0.01\kMpc$ ({\em lower panels}) of the matter bispectrum. The left panels show the full prediction at one-loop assuming $\fNL=100$ ({\em black, continuous line}), together with the tree-level components $B_{111}$ ({\em short dashed, red}) and $B_{112}^I$ ({\em long-dashed, blue}) and the one-loop corrections present for Gaussian initial conditions ({\em dot-dashed, blue}) and those depending instead on the initial bispectrum and trispectrum ({\em dotted, red}). The central panels show the individual terms of the one-loop corrections for Gaussian initial conditions, while the right panels show the individual components of the one-loop corrections present only for non-Gaussian initial conditions. }
\label{fig:bscomp}
\end{figure*}
In \fig{fig:bscomp} we show the different components in PT to the equilateral configurations ({\em upper panels}) and to the squeezed configurations ({\em lower panels}) of the matter bispectrum, respectively $B(k,k,k)$ and $B(k,k,\Delta k)$ with $\Delta k \simeq 0.012\kMpc $ as a function of $k$. In the left panels, we compare the tree-level contributions $B_{111}$ with $\fNL=100$ ({\em short-dashed, red lines}) and $B_{112}^{I}$ ({\em long-dashed, blue lines}) to the sum of the one-loop corrections $B_{222}+B_{123}^I+B_{123}^{II}+B_{114}$ ({\em dot-dashed, blue lines}) present for Gaussian initial conditions
and to the sum of the one-loop corrections due to primordial non-Gaussianity, $B_{112}^{II}+B_{122}^I+B_{122}^{II}+B_{113}^I+B_{113}^{II}$, with $\fNL=100$ ({\em dotted, red lines}). The central and right panels compare these sums of one-loop corrections to their individual contributions. For the ``Gaussian'' piece ({\em central panels}), notice that we plot $-B_{123}^{II}$ and $-B_{114}$, implying that the overall one-loop correction is the result of a number of cancellations similarly to those occurring for the one-loop corrections to the matter power spectrum. Analogous considerations also apply to the ``non-Gaussian'' one-loop corrections ({\em right panels}) where such cancellations strongly depend on the triangular configuration. 

To conclude the section, we note that the ``order'' of each correction in the perturbative expansion is defined in terms of the power of linear {\em matter} density field, $\d^{(1)}$.  An alternative convention could be given by counting the powers of the Gaussian contribution to the curvature perturbations, that is $\phi$ in \eqn{eq:phiNG}. Our choice is motivated by the standard use in the large-scale structure literature, and by the fact that to the $n$-th perturbative order corresponds the well defined redshift dependence $D^n(z)$. On the other hand, we can keep track of the expansion in $\phi$ in terms of the nonlinear parameter $\fNL$.

\section{Simulations}
\label{sec:sims}

We utilize a series of large 1024$^3$ N-body simulations of the $\Lambda$CDM cosmology seeded with  Gaussian and non-Gaussian initial conditions \citep{DesjacquesSeljakIliev2009}. The box size is 1600$\Mpc$ with a force resolution of 0.04 times the mean inter-particle distance. The (dimensionless) power spectrum of the Gaussian part $\phi(\xv)$ of the Bardeen potential is the usual power-law $\Delta_\phi^2(k)\equiv  k^3 P_\phi(k)/(2\pi^2)=A_\phi (k/k_0)^{n_s-1}$.  The non-Gaussianity  is of the local form described above. We adopt the standard (CMB) convention in which $\Phi(\xv)$ is primordial, and not extrapolated to present epoch. We assume $h=0.7$, $\Omega_{\rm m}=0.279$, $\Omega_{\rm b}=0.0462$, $n_s=0.96$, and a normalization of the Gaussian curvature perturbations $A_\phi=7.96\times 10^{-10}$ at the pivot point $k_0=0.02$Mpc$^{-1}$, close to the best-fitting values inferred from CMB measurements \citep{KomatsuEtal2009B}. This yields a density fluctuations amplitude $\sigma_8\simeq 0.81$ when the initial conditions are Gaussian. Eight sets of three simulations, each of which has $\fNL=0,\pm 100$, were run with the N-body code {\scshape gadget} \citep{Springel2005}. The same Gaussian random seed field $\phi$ is employed in each set of runs so as to minimize the sampling variance. The initial particle distribution is generated at redshift $z_i=99$ using the Zel'dovich approximation \citep{Zeldovich1970}.

\subsection{Bispectrum estimator and triangle bins}
\label{sec:est}

Let us now introduce the bispectrum estimator $\hat{B}(\kall)$ used in the analysis of the N-body simulations. For a cubic box of volume $V$, this is given by \citep{ScoccimarroEtal1998}
\beq
\label{Best}
\hat{B} \equiv \frac{V_f}{V_{B}}\int_{k_1}\!\!\!\!d^3 q_1\!\!\int_{k_2}\!\!\!\!d^3 q_2\!\!\int_{k_3}\!\!\!\!d^3 q_3\,\delta_D(\qv_{123})\, \d_{\qv_1}\,\d_{\qv_2}\,\d_{\qv_3}\,,
\eeq
where $V_f\equiv k_f^3=(2\pi)^3/V$ is the volume of the fundamental cell and where each integration is defined over the bin $q_i\in[k_i-\D k/2,k_i+\D k/2]$ centered at $k_i$ and of size $\D k$ equal to a multiple of the fundamental frequency $k_f$. In our case we assume a bin size $\D k=3 k_f$. The Dirac delta function $\d_D(\qv_{123})$ ensures that the wavenumbers $\qv_1$, $\qv_2$ and $\qv_3$ indeed form a closed triangle, as imposed by translational invariance. The normalization factor $V_B(\kall)$ given by
\bea
V_B(\kall) & \equiv & \int_{k_1} \!\!\!\! d^3 q_1\int_{k_2} \!\!\!\! d^3 q_2 \int_{k_3} \!\!\!\! d^3 q_3 \,\delta_D(\qv_{123})\nonumber\\
& \simeq & 
8\pi^2\ k_1 k_2 k_3\ \D k^3
\eea
represents the number of {\it fundamental} triangular configurations (labelled by the triplet $\qv_1$, $\qv_2$ and $\qv_3$) that belong to the triangular configuration {\it bin} defined by the triangle sizes $k_1$, $k_2$ and $k_3$ with uncertainty $\D k$. In order to better interpret the simulation results, we provide as well the expression for the variance of the bispectrum associated with this estimator. At leading order, the variance reads as \citep{ScoccimarroEtal1998}
\beq\label{eq:Bvariance}
\D B^2(\kall)=s_B\frac{V_f}{V_B}P(k_1)P(k_2)P(k_3),
\eeq
where the symmetry factor $s_B(\kall)=6$, $2$ or $1$ for equilateral, isosceles or scalene configurations. This expression neglects further corrections depending on the matter bispectrum, trispectrum and six-point functions that are responsible for correlations between different configurations \citep[see][]{SefusattiEtal2006}. 

When comparing the measured bispectrum configurations to the theoretical predictions in perturbation theory, one should be careful to properly account for the effect of the finite size of the triangle bins. As explained above, each configuration is defined in terms of the sides of the triangle with $k_i$ being the central value and $\Delta k$ the uncertainty. Since we are assuming $\Delta k=3 k_f$, a typically large number of ``fundamental'' triangles fall into each triangle bin. For instance, it is easy to see that, in the case of equilateral configurations, the bin defined by the central value $k$ will include equilateral triangles of side $q=k-k_f$ or $q=k+k_f$ just as well as nearly-equilateral triangles with different sides still belonging to the $k$-bin. What is important here is the fact that, in the case of equilateral configurations, we will have slightly {\em more} triangles of size larger than the fundamental equilateral triangle with side $q=k$, than triangles of smaller size. This simply follows from the larger number of modes at larger $q$. 

The correct approach consists in computing the raw PT prediction $B^{PT}(q_1,q_2,q_3)$ {\em and average it over the triangle bin} defined by $k_1$, $k_2$, $k_3$ and $\Delta k$, that is
\bea
B^{th}(\kall)& = &\frac{1}{V_{B}}\int_{k_1}\!\!\!\!d^3 q_1\int_{k_2}\!\!\!\!d^3 q_2\int_{k_3}\!\!\!\!d^3 q_3 \;\delta_D(\qv_{123})\times
\nonumber\\
& & B^{PT}(q_1,q_2,q_3)\,,
\eea
where $B^{th}$ is the value to be compared with the measurements. This is, however, computationally challenging especially in the case of the one-loop corrections to the bispectrum, which usually involve three-dimensional integrations. An alternative solution, less rigorous yet reasonable given the uncertainties of our measurements, consists in defining the following {\em effective} values $\tilde{k}_i$ for the wavenumbers $k_i$ characterizing the triangle,
\beq
\tilde{k}_i  =  \frac{1}{V_{B}}\int_{k_1}\!\!\!\!d^3 q_1\int_{k_2}\!\!\!\!d^3 q_2\int_{k_3}\!\!\!\!d^3 q_3 \;\delta_D(\qv_{123})~ q_i\,,
\eeq
so that the theoretical prediction which the binned measurements of the bispectrum must be compared to is
\beq
B^{th}(\kall) =  B^{PT}(\tilde{k}_1,\tilde{k}_2,\tilde{k}_3)\,.
\eeq
This procedure improves significantly the agreement between theory and simulations, particularly for ``squeezed'' configurations where $k_1\ll k_2\simeq k_3$. 

Here and henceforth, all theoretical predictions will be computed in terms of the effective triangle $\tilde{k}_1$, $\tilde{k}_2$ and $\tilde{k}_3$ as defined above. Furthermore, when the bispectrum is expressed as a function of the angle $\theta$ between two of the three wavemodes, it is convenient to introduce an {\it effective} angle $\tilde{\theta}$ given by
\bea\label{eq:thetaeff}
\cos\tilde{\theta}(k_1,k_2;k_3) & = & \frac{1}{V_{B}}\int_{k_1}\!\!\!\!d^3 q_1\int_{k_2}\!\!\!\!d^3 q_2\int_{k_3}\!\!\!\!d^3 q_3 \;\delta_D(\qv_{123})\times
\nonumber\\ & & 
 \cos\theta(q_1,q_2;q_3)\,,
\eea
where $\theta(q_1,q_2;q_3)$ is the angle between the vectors $\qv_1$ and $\qv_2$. This expression defines the effective angle as a weighted average of the angles corresponding to the "fundamental" triangles falling in a given bin. These are limited by the triangle inequalities $q_3\le q_1+q_2$ and $q_3\ge |q_1-q_2|$. In the figures, the quantities measured in the N-body simulations will be plotted as a function of $\tilde{\theta}$, while the theoretical expectations will be the raw PT predictions.

\subsection{The power spectrum}

To facilitate the comparison between different statistics and help interpreting the bispectrum measurements of the next section, we will first present measurements of the matter power spectrum, highlighting the effects due to primordial non-Gaussianity and their description in PT. Similar results can be found in \citet{DesjacquesSeljakIliev2009, GrossiEtal2008, PillepichPorcianiHahn2010, BartoloEtal2010}. 

\begin{figure*}
{\includegraphics[width=0.95\textwidth]{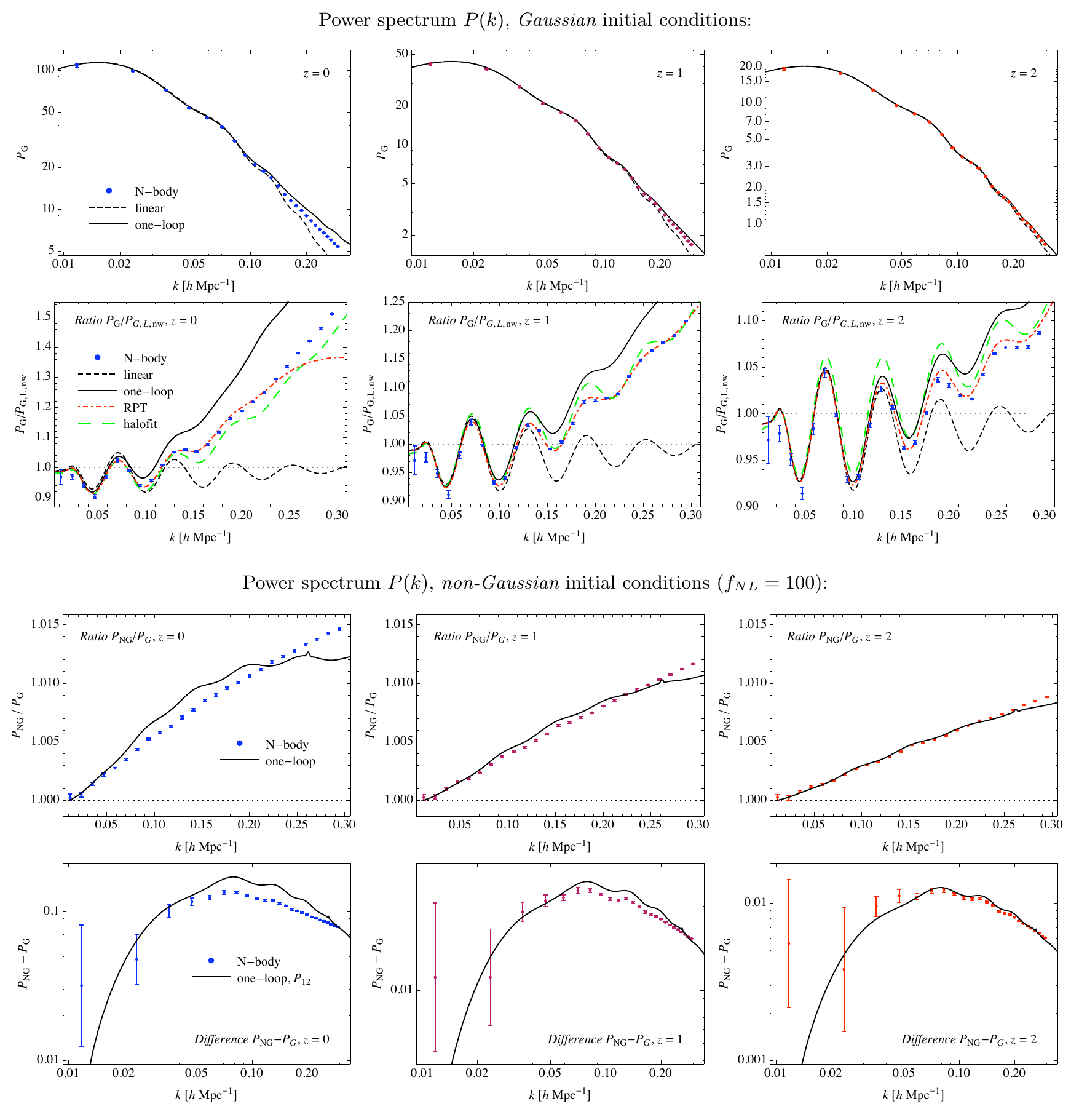}}
\caption{Measurements of the matter power spectrum, $P(k)$, as a function of $k$. We show, from top to bottom, the power spectrum $B$ ({\em first row}) and its ratio to the no-wiggle, linear prediction ({\em second row}) for Gaussian initial conditions, the ratio $P(\fNL=+100) / P(\fNL=0)$ ({\em third row}) and the difference $P(\fNL=+100) - P(\fNL=0)$ ({\em last row}). Different columns correspond to redshifts $z=0$, $1$ and $2$. Short-dashed, black line indicate the tree-level PT predictions while continuous, black lines the one-loop ones. In addition, on the second row we include the RPT prediction of \citet{CrocceScoccimarro2006A} (at the two-loop approximation, {\em dot-dashed, red line}) and the prediction from the code \texttt{halofit} of \citet{SmithEtal2003} ({\em long-dashed, green line}).}
\label{fig:psmG}
\end{figure*}
In the two upper rows of \fig{fig:psmG} we show the matter power spectrum measured in simulations of Gaussian initial conditions, as well as the linear ({\em dashed lines}) and one-loop ({\em continuous lines}) predictions in PT. In addition, in the second row, displaying the ratio of the Gaussian power spectrum with a smooth (\ie no-wiggles) linear power spectrum, we show the nonlinear power spectrum obtained with the \texttt{halofit} code of \citet{SmithEtal2003} ({\em thin, green line}) and the predictions in Renormalized Perturbation Theory (RPT) \citep{CrocceScoccimarro2006A, CrocceScoccimarro2006B, CrocceScoccimarro2008}. The various columns correspond, from left to right, to the redshift $z=0$, $1$ and $2$, respectively. The well-known failure of one-loop PT to describe the matter power spectrum at mildly nonlinear scales and low redshift is quite apparent \citep[see][for a recent comparison with simulations and for a comparison at high redshift, respectively]{CrocceScoccimarro2008, JeongKomatsu2006}. The \texttt{halofit} prediction is significantly better at low redshift, while it shows a discrepancy of the same order at $z=2$. On the other hand, the RPT prescription provides very good predictions (within 1\%) up to $0.23\kMpc$ at redshift zero, and over the whole range we consider at redshift $z=1$ and $2$. The slight discrepancy at $z=2$, of the order of $0.5\%$, not present at $z=1$, might be perhaps be explained in terms of transients from the initial conditions \citep{Scoccimarro1998, CroccePueblasScoccimarro2006}, despite the relatively high redshift ($z=99$) assumed for the simulations.  

In the third row, we show the ratio between the matter power spectrum extracted from the $\fNL=100$ and Gaussian simulations. The plots for $z=0$ and $z=2$ reproduce Fig.~3 in \citet{DesjacquesSeljakIliev2009}. Finally, the last row shows the {\em difference} between the two cases, \ie $P(k;\fNL=100)-P(k;\fNL=0)$. In all these plots, the ratio and the difference measured in the simulations are computed for each realization and then averaged over the eight realizations available. At redshift zero, the one-loop correction $P_{12}$ reproduces qualitatively the effect due to primordial non-Gaussianity, but it breaks down at relatively large scales, $k\sim 0.2\kMpc$, maybe suggesting the need for higher order corrections. An extension of perturbation theory such as the time-renormalization group approach \citep{MatarresePietroni2007, Pietroni2008} seems to improve only in part the agreement between theory and simulations beyond this scale \citep[see Fig.~4 in][]{BartoloEtal2010}. 

In each realization of the initial conditions with $\fNL=0$, $\pm 100$, we also measured the combination $[P(k;\fNL=+100)+P(k;\fNL=-100)-2 P(k;\fNL=0)]/2$. In the PT framework, the result is expected to be the sum of all the corrections depending on even powers of $\fNL$. At the lowest order however, these are given by two-loop contributions which we ignore in this work. Nevertheless, we find that in the range of scale considered here and for $\fNL=100$, such terms represent an effect of the order of $10^{-4}$ relative to the power spectrum for Gaussian initial conditions.

\section{Results}
\label{sec:results}

We now present the measurements of the matter bispectrum with Gaussian and non-Gaussian initial conditions together with one-loop PT predictions. In the figures, we will often denote these quantities as $B_{G}$ and $B_{NG}$, where the ``G'' and ``NG'' subscripts refer to the initial conditions. In the Gaussian case moreover, we will also perform a comparison between the measurements and the fitting formula of \citet{ScoccimarroCouchman2001}. 
 
To assess the agreement between PT and N-body measurements as a function of scale and triangle shape, we will consider five sets of configurations. We will present results as a function of $k$ for equilateral configurations $B(k,k,k)$, isosceles configurations $B(2k,2k,k)$ as well as increasingly ``squeezed'' configurations $B(k,k,\D k)$ with fixed $\Delta k$. To further explore the shape dependence, we will also show the result of measuring the matter bispectrum for two sets of generic configurations for which the magnitude of two sides of the triangle ($k_1$ and $k_2$) is fixed while the angle $\theta$ between them is varied.

For each of these sets, in each of the following figures, the upper two panels show measurements of the matter bispectrum $B$, or the {\em reduced} bispectrum $Q$, see \eqn{eq:qbs} below, for {\em Gaussian} initial conditions, as well as the ratio to the corresponding tree-level expression in PT where the acoustic oscillations are removed by means of the smooth transfer function of \citet{EisensteinHu1998}. Recall that there is no ``linear'' matter bispectrum for Gaussian initial conditions (but there is an initial bispectrum in the presence of primordial non-Gaussianity). For sake of comparison, we take the tree-level prediction as a reference since it is most directly related to the linear bispectrum, which is generically $B^{tree}\sim P_L^2$. 

The last three rows in the plots focus on the effect of primordial non-Gaussianity. We show, in particular, the ratio 
\bdm
B(\fNL=100)/B(\fNL=0)\quad \textrm{({\em third~row})}\,,
\edm
the difference 
\bdm
B(\fNL=100)-B(\fNL=0)\quad \textrm{({\em fourth row})}
\edm
with respect to the Gaussian case, and the combination
\bdm
[B(\fNL=+100)+B(\fNL=-100)
\edm\vspace{-7mm}
\bdm
-2~B(\fNL=0)]/2 \quad\textrm{({\em last row})}
\edm
to highlight the effects proportional to $\fNL^2$. In all cases, the N-body results indicate the mean over eight realizations of the specific combination (ratio, difference, etc.) performed with Gaussian and non-Gaussian initial conditions drawn from the same random seed field $\phi$ (see section~\ref{sec:sims}). In this way, we can study the effect of non-Gaussianity without the additional sampling variance affecting, for instance, the difference $B_{NG}-B_G$ obtained as the {\em difference} between the {\em mean} $B_{NG}$ and the {\em mean} $B_G$ over the eight realizations. Finally, the three columns correspond to the results at redshift $z=0$, $1$ and $2$. In all the plots, the numerical results are compared to the tree-level ({\em short-dashed, black lines}) and one-loop predictions ({\em continuous, black lines}) in PT.    

\begin{figure*}
{\includegraphics[width=0.95\textwidth]{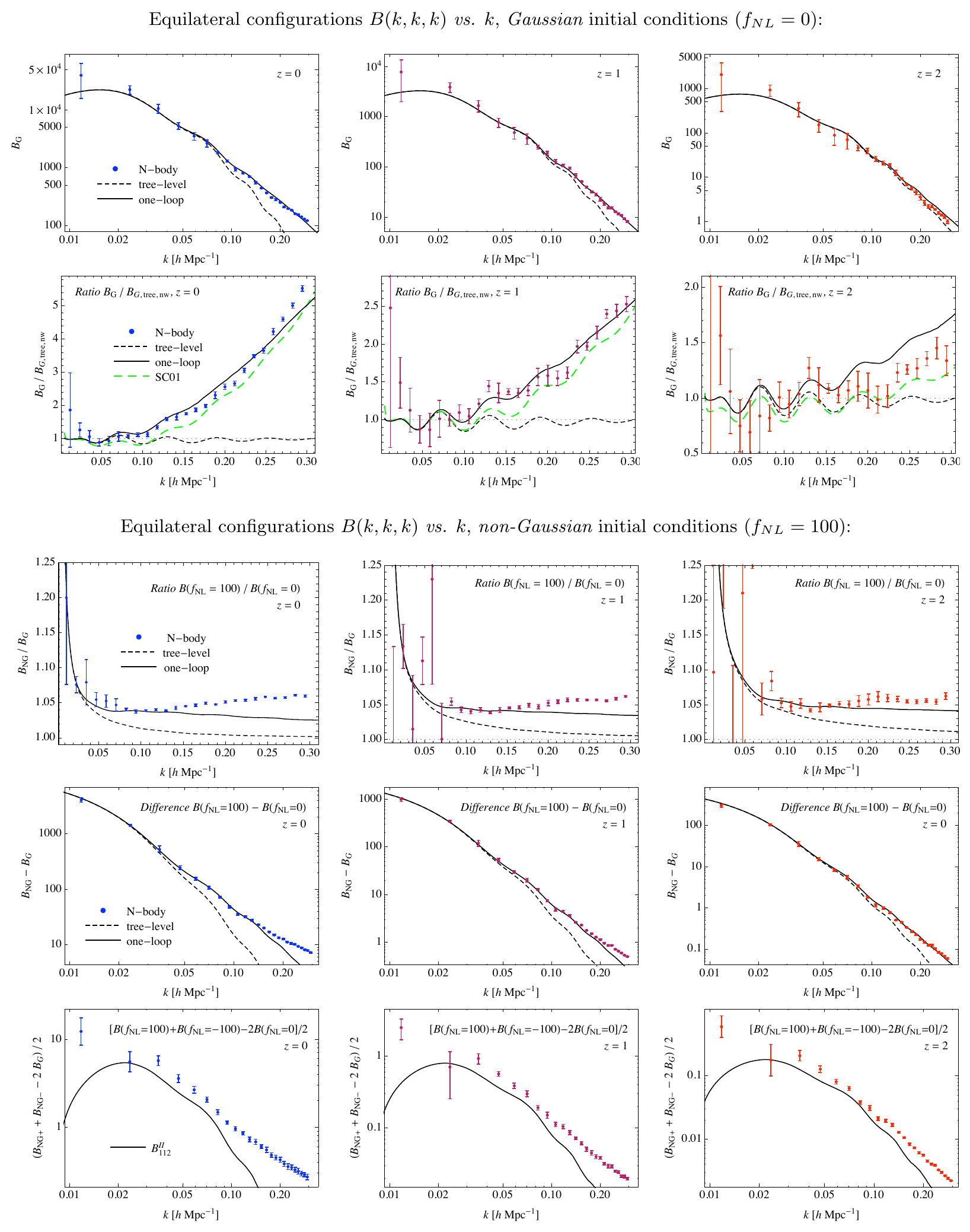}}
\caption{Measurements of the {\bf equilateral configurations} of the matter bispectrum, $B(k,k,k)$, as a function of $k$. We show, from top to bottom, the matter bispectrum $B$ ({\em first row}) and its ratio to the no-wiggle tree-level prediction ({\em second row}) for Gaussian initial conditions, the ratio $B(\fNL=+100) / B(\fNL=0)$ ({\em third row}), the difference $B(\fNL=+100) - B(\fNL=0)$ ({\em fourth row}) and the combination $[B(\fNL=+100)+B(\fNL=-100)-2B(\fNL=0)]/2$ ({\em last row}). Different columns correspond to redshifts $z=0$, $1$ and $2$. Short-dashed, black line indicate the tree-level PT predictions while continuous, black lines the one-loop ones. In addition, on the second row we include the fitting formula of \citet{ScoccimarroCouchman2001} ({\em long-dashed, green lines}). }
\label{fig:bsmEq}
\end{figure*}
In \fig{fig:bsmEq}, we show the matter bispectrum $B(k,k,k)$ for equilateral configurations. As can be seen, non-linearities are particularly severe, consisting in a almost $\sim 300\%$ correction relative to the tree-level prediction for $k\simeq 0.2\kMpc$ and $z=0$ for instance. The bispectrum measured from a total simulation volume of $\sim 33\cGpc$ presents errors of the order of 10\% at this scale for equilateral configurations. Notice that this specific triangle shape suffers, unlike other configurations close in shape and scale, from a relatively large variance (up to a factor of six). This effect originates partly from the symmetry factor $s_B$ in \eqn{eq:Bvariance}, and from the large contribution of higher-order correlation functions to the bispectrum variance. 

The one-loop prediction appears to be well within our errors up to $k\sim 0.15\kMpc$ and describes reasonably well the behavior at smaller scales. For $k\lesssim 0.15\kMpc$, the one-loop prediction behaves better than the fitting formula of \citet{ScoccimarroCouchman2001} (in the plots SC01), which under-predicts the data points at mildly non-linear scales. This $\sim 20\%$ discrepancy, unsurprising given the size of the simulation box used for the fit ($240\Mpc$), has already been noted by \citet{PanColesSzapudi2007}. It should be remarked that the SC01 formula aimed at describing the nonlinear bispectrum at smaller scales, particulalrly for weak lensing applications, and did not addressed specifically the issue of the acoustic features. \citet{PanColesSzapudi2007} also proposed a phenomenological model for the matter bispectrum based on a rescaling argument similar to the one explored in \citet{HamiltonEtal1991, PeacockDodds1996}. We also compared this prescription to our measurements and find that it agrees better than the fitting function of \citet{ScoccimarroCouchman2001}. However, the rescaling induces an large and unphysical shift in the acoustic oscillations that should be properly accounted for \citep[in][comparisons are shown with simulations of featureless matter power spectra]{PanColesSzapudi2007}. 
  
The third row of \fig{fig:bsmEq} shows the effect of primordial non-Gaussianity in terms of the ratio $B(\fNL=100)/B(\fNL=0)$. It is interesting to notice that the additional non-linear contributions due to non-Gaussian initial conditions correspond, for these set of configurations, to a $\sim 5\%$ correction regardless of redshift. In fact, the contribution of the initial bispectrum $B_0$ to this effect is already subdominant at $k\sim 0.1\kMpc$ and $z=0$, while one-loop corrections themselves fail to account for it at slightly smaller scales. This is also apparent in the difference $B(\fNL=100) - B(\fNL=0)$ which, in the PT picture, arises from the one-loop contributions depending on the initial bispectrum and trispectrum. At redshift zero, these provide an accurate description of $B(\fNL=100) - B(\fNL=0)$ up to $k~0.15\kMpc$.   

Finally, in the last row we compare the combination $[B(\fNL=+100)+B(\fNL=-100)-2~B(\fNL=0)]/2$ to $B_{112}^{II}$ which, in the one-loop approximation, is the sole term depending on the initial trispectrum and, therefore, on $\fNL^2$. This term appears to underestimate by about $50\%$ (at best) the simulation results. One should nonetheless keep in mind these contributions represent a $0.1\%$ correction to the matter bispectrum. 

\begin{figure*}
{\includegraphics[width=0.95\textwidth]{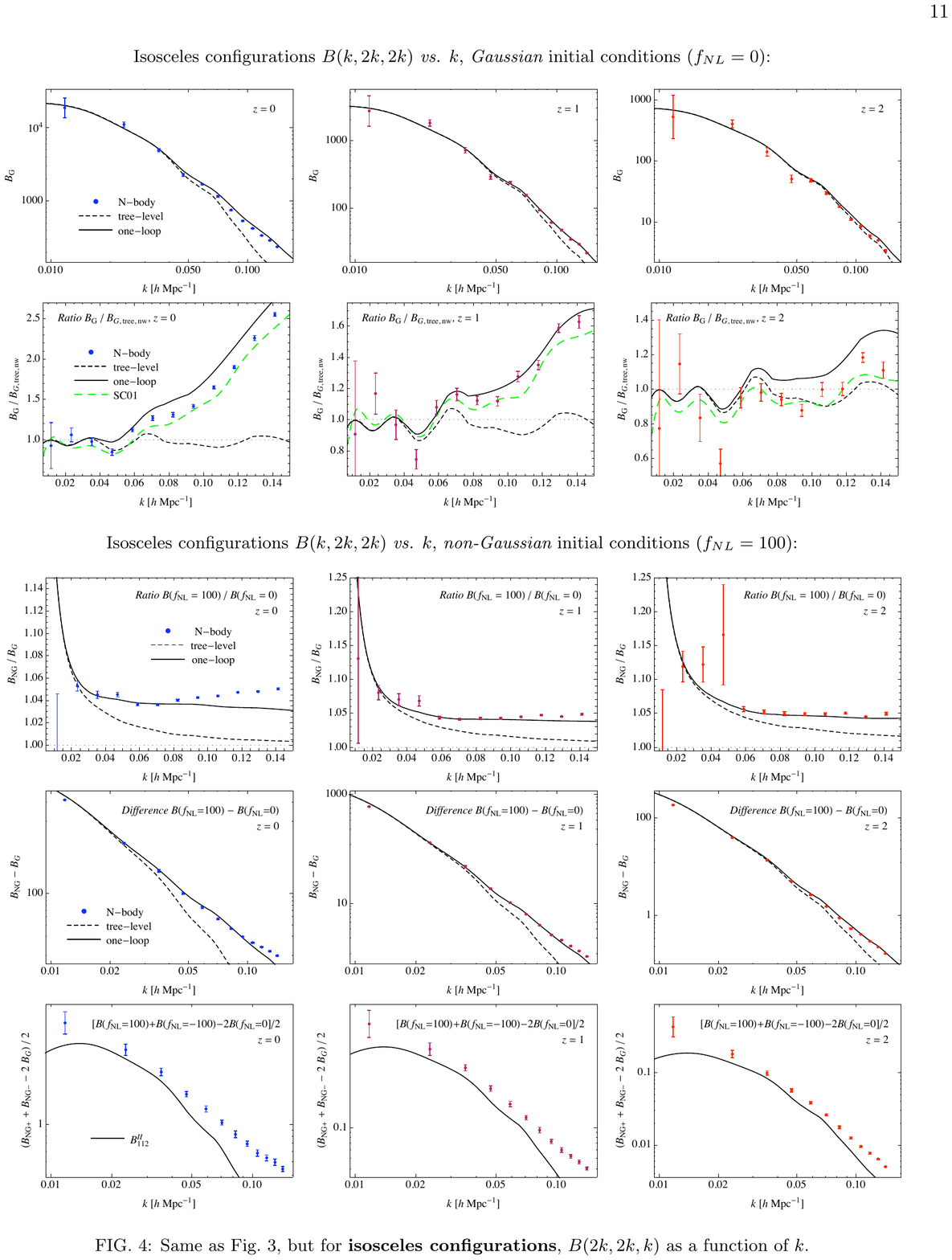}}
\caption{Same as \fig{fig:bsmEq}, but for {\bf isosceles configurations}, $B(2k,2k,k)$ as a function of $k$.}
\label{fig:bsmIso}
\end{figure*}
In \fig{fig:bsmIso}, we show the matter bispectrum for the isosceles configurations, $B(2k,2k,k)$, as a function of $k$. The shape of the triangle is unchanged while its size is rescaled. In this series of plots, the relatively smaller variance (with respect to that of the equilateral shape) expected from the discussion above is quite apparent. The error on the mean is of the order of $2$-$3\%$ for most of the isosceles configurations considered. These small errors allow a more accurate comparison of the measurements with PT predictions. Note that, while each triangle now involves two different scales $k$ and $2k$, the results are shown as a function of the smaller one ($k$) solely. For Gaussian initial conditions, the one-loop predictions systematically overestimates the data points by more than $10\%$ at $z=0$, but the agreement substantially improves at higher redshift. By contrast, the accuracy of the fitting formula of SC01 is reasonably good for all the scales and redshifts considered. As for the effect of primordial non-Gaussianity, considerations similar to those made for equilateral configurations also hold for the isosceles shape. 

\begin{figure*}
{\includegraphics[width=0.95\textwidth]{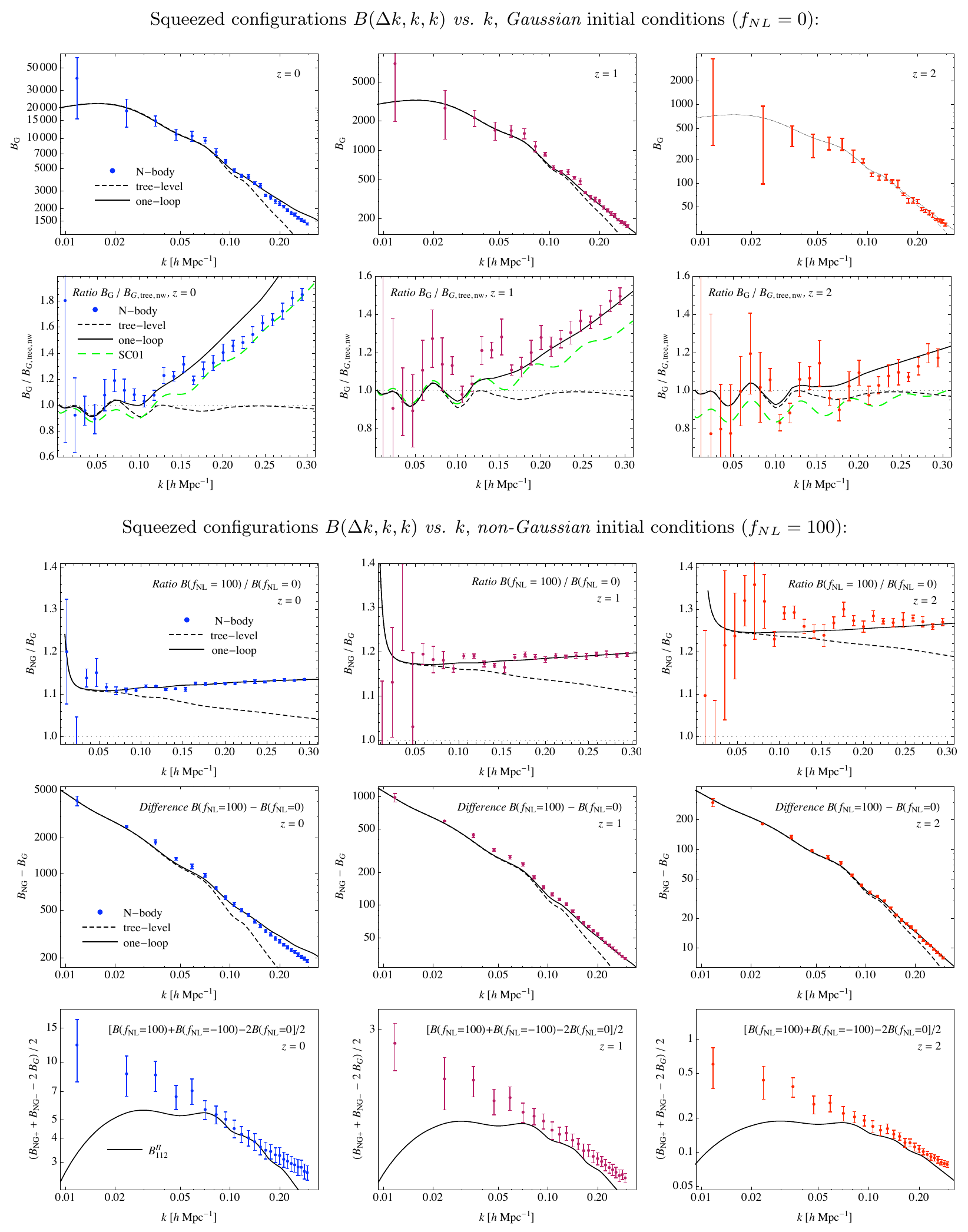}}
\caption{Same as \fig{fig:bsmEq}, but for {\bf squeezed configurations}, $B(\D k,k,k)$, with $\D k = 3k_f\simeq 0.012\kMpc$ as a function of $k$.}
\label{fig:bsmSq}
\end{figure*}
In \fig{fig:bsmSq} we compute $B(k,k,\Delta k)$ on triangles one side of which is held fixed to the smallest available $k$-bin $\D k$ while the other two are equal and varying. $k$ is increased smoothly such that this configuration, which represents the coupling between the scales $k$ and $\D k$, asymptote to the ``squeezed'' triangle shape. The errors on this {\em highly correlated} set of configurations are dominated by the large variance of the small-scale mode $\D k$ and are typically slightly larger than 10\%. Still, the one-loop approximation for the Gaussian case breaks down already around $k=0.15\kMpc$ at redshift zero. The SC01 formula shows instead the same discrepancy noted above around $k\sim 0.1\kMpc$ while it provides a better fit to the data at larger $k$. At higher redshift however, perturbation theory fares better than the fitting formula. The limitation of the one-loop prediction for the Gaussian case is also apparent in the corrections due to non-Gaussianity. However, the theoretical prediction for the {\em ratio} $B_{NG}/B_G$ is in remarkable agreement with the data (third row of \fig{fig:bsmSq}). Note that the large-$k$ limit in this set of configurations does not correspond to the more common ''squeezed'' limit, obtained fixing two sides of the triangle and reducing the third one, so we do not expect an increase in the non-Gaussian component, since at larger $k$ we are probing smaller scales and the suppression due to the transfer function is larger. Nevertheless, the non-Gaussian corrections are relatively large for this set of configuration, ranging from $10$ to $30\%$ and growing with redshift. This triangle shape is, among those we consider, the most directly comparable to the measurements of \citet{NishimichiEtal2009} at redshift $z=0.5$ and, in particular to the central panels of their Fig. 3. Our errors are consistent with theirs, and the agreement between our data points and tree-level PT at $z=0.5$ is also reasonable. 

In the last two figures, we consider generic scalene triangles for which the length of two sides $k_1$ and $k_2$ is held fixed while the angle $\theta$ between them (and, therefore, the length of the third side) is varied. Such a set of triangular configurations is useful to illustrate the shape dependence of the bispectrum as it includes collapsed, flattened and almost equilateral triangles depending on the choice of $k_1$ and $k_2$. To further isolate the shape dependence of the matter bispectrum from its scale dependence, it is convenient to introduce the {\em reduced} bispectrum $Q(\kall)$ defined as
\beq\label{eq:qbs}
Q\equiv\frac{B(\kall)}{P(k_1)P(k_2)+P(k_1)P(k_3)+P(k_2)P(k_3)}.
\eeq  
In the following two figure, we will indeed show the reduced bispectrum in the first rows instead of the bispectrum itself. Notice that the one-loop predictions for the reduced bispectrum are computed from a proper expansion of the denominator in terms of the one-loop expression for the power spectrum \citep[see][for details]{Sefusatti2009}. The quantities shown in the other rows are the same as before. A second difference with the previous plots is the fact that the data points are plotted as a function of the effective angle $\theta$ defined in \eqn{eq:thetaeff} (see section~\ref{sec:est}). 

\begin{figure*}
{\includegraphics[width=0.95\textwidth]{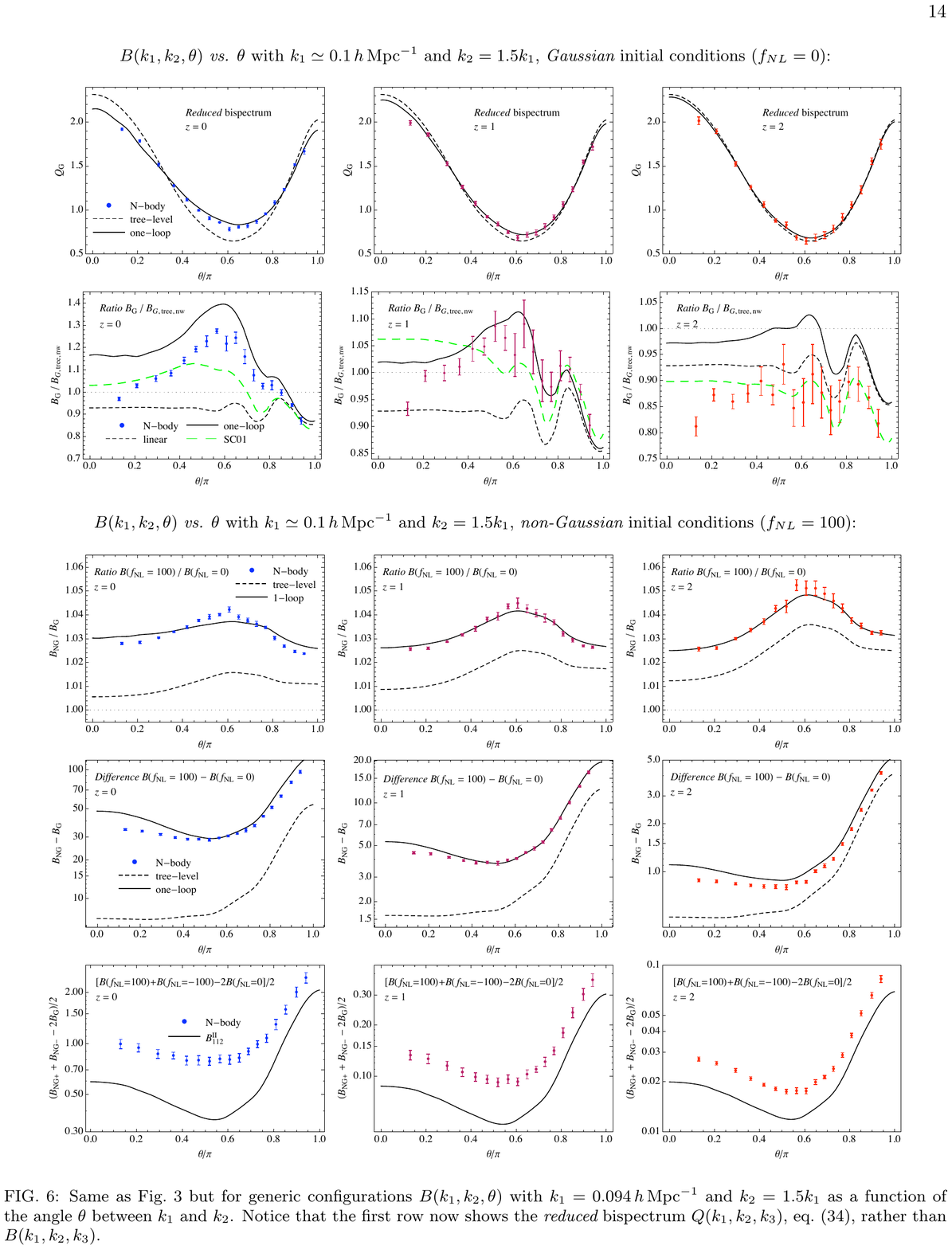}}
\caption{Same as \fig{fig:bsmEq} but for generic configurations $B(k_1,k_2,\theta)$ with $k_1=0.094\kMpc$ and $k_2=1.5~k_1$ as a function of the angle $\theta$ between $k_1$ and $k_2$. Notice that the first row now shows the {\em reduced} bispectrum $Q(\kall)$, \eqn{eq:qbs}, rather than $B(\kall)$.}
\label{fig:qbsmA}
\end{figure*}
In figure~\ref{fig:qbsmA}, we consider the specific case $k_1=0.094\kMpc$ and $k_2=1.5~k_1$. For these configurations, $\theta\lesssim 0.6~\pi$ implies that all three sides are larger than $0.1\kMpc$. On these scales, the agreement of the one-loop predictions with the measurements at $z=0$ is poor, as is evident from the first plots on the second row. Errors on the bispectrum mean are typically of the order of $3\%$. At redshift $z\gtrsim1$ however, the theoretical predictions fall within the errors. Rather puzzling is, however, the relatively poor agreement at $z=2$, in fact present already in the previous plots and perhaps related to small descrepancy between RPT predictions and simulations in the power spectrum case. The prediction for the relative effect of primordial non-Gaussianity, which is about $3\%$ at all redshift, is in good agreement with the data regardless of the triangle shape ({\em third row}). The apparent bump shown in these plots results from the low values of the ``Gaussian'' bispectrum for nearly equilateral triangles evident from the plots in the first row, rather then a non-Gaussian feature. Instead, the larger non-Gaussian signal expected for triangles approaching the squeezed limit is observable in the ``difference'' plots on fourth row for $\theta \simeq \pi$\footnote{When the ``direction'' of the three wavevectors with sum equal to zero, \ie $\kv_1+\kv_1+\kv_3=0$, is taken into account it easy to see that the ''squeezed'' limit is obtained for $\theta\rightarrow\pi$, rather then $\theta\rightarrow 0$ as one might na\"ively think just considering the the triangle defined by the wavenumber magnitudes alone.}. Notably, the same feature appears also in the component $B_{112}^{II}$ dependent on the initial trispectrum $T_0$ ({\em last row}).   

\begin{figure*}
{\includegraphics[width=0.95\textwidth]{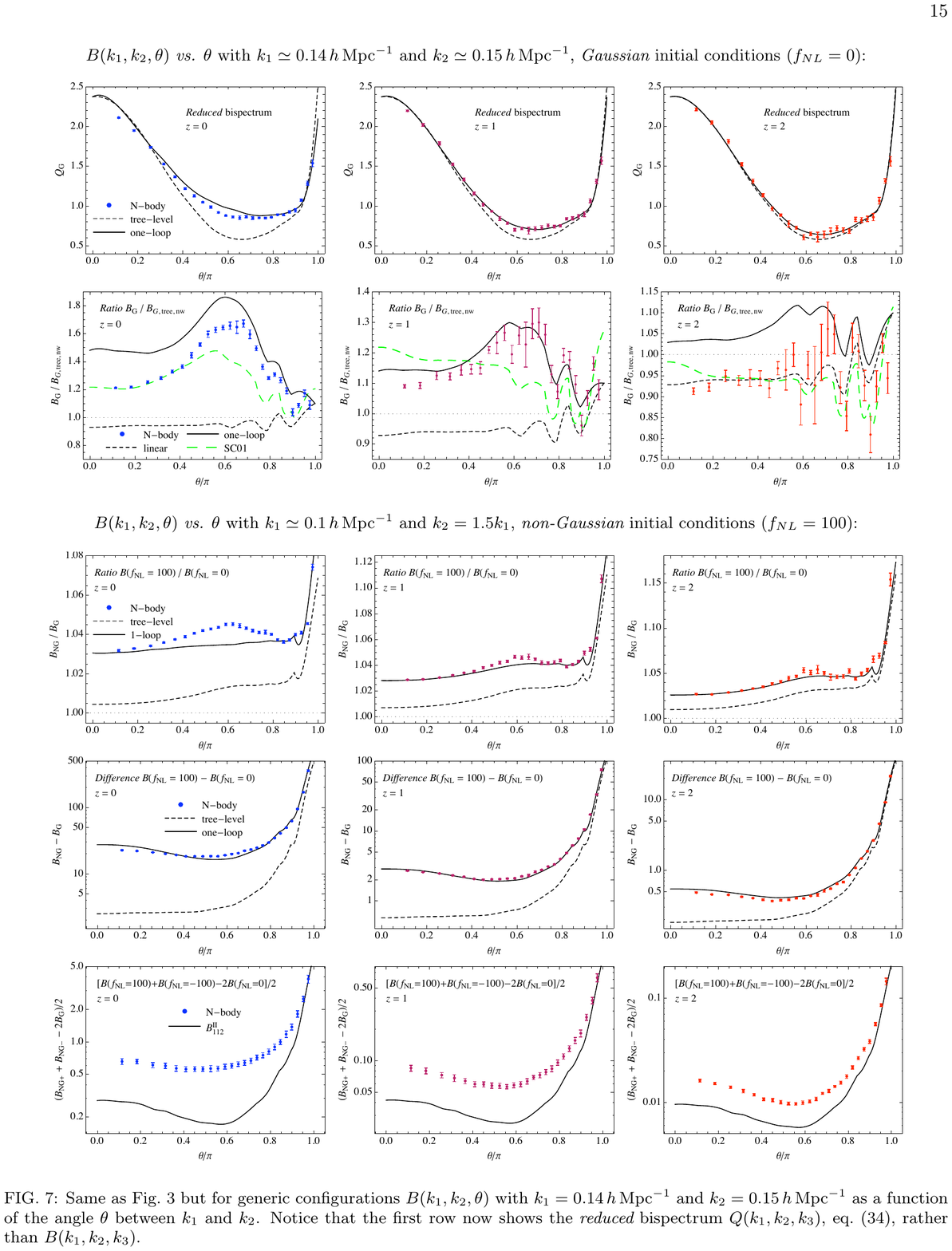}}
\caption{Same as \fig{fig:bsmEq} but for generic configurations $B(k_1,k_2,\theta)$ with $k_1=0.14\kMpc$ and $k_2=0.15\kMpc$ as a function of the angle $\theta$ between $k_1$ and $k_2$. Notice that the first row now shows the {\em reduced} bispectrum $Q(\kall)$, \eqn{eq:qbs}, rather than $B(\kall)$.}
\label{fig:qbsmB}
\end{figure*}
Similar results are found for a second set of triangles where the two sides are now much closer in size, $k_1=0.14\kMpc$ and $k_2=0.15\kMpc$ (see figure~\ref{fig:qbsmB}). In this case however, the configurations are very close to equilateral for $\theta\simeq 0.6~\pi$. As a result, we observe at $z=0$ the same discrepancy between PT and simulations than that seen in figure~\ref{fig:bsmEq} at small scales. This disagreement is also apparent in the plot of the reduced bispectrum. The non-Gaussian correction typically is of the order of $3\%$, but it increases significantly in the squeezed limit $\theta\rightarrow\pi$. This behavior of the linear and nonlinear components due to primordial non-Gaussianity is also evident in the fourth row showing the difference $B_{NG}-B_G$.

\section{Conclusions}
\label{sec:conclusions}

A rather surprising effect of local primordial non-Gaussianity on the large scale clustering properties of biased objects has been observed in various numerical studies over the last years \citep{DalalEtal2008, DesjacquesSeljakIliev2009, PillepichPorcianiHahn2010, GrossiEtal2009}. These results attracted a great deal of attention as they showed that measurements of the power spectrum of galaxies and quasars from current data sets can lead to constraints on the local non-Gaussian parameter $\fNL$ comparable to those of CMB observations \citep{SlosarEtal2008, DesjacquesSeljak2010B}. Previous work assumed that the main effect of primordial non-Gaussianity is limited to an extra contribution to the matter and galaxy bispectrum. Still, even under such incorrect but ``conservative'' assumption, it has been shown that future large-volume redshift surveys will reach a sensitivity to a non-zero $\fNL$ comparable or better than the CMB bispectrum \citep{ScoccimarroSefusattiZaldarriaga2004, SefusattiKomatsu2007}. The inclusion of the non-Gaussian bias in the analysis of the galaxy bispectrum or, better, in a combined analysis of the power spectrum and bispectrum, is desirable to reliably assess the potentiality of forthcoming surveys of the large scale structure.  

As a first step in this direction, we have measured the matter bispectrum for the main classes of triangle shape using a set of large-volume N-body simulations seeded with Gaussian and non-Gaussian initial conditions of the local type. We focused on mildly nonlinear scales, $0.02\lesssim k\lesssim 0.3\kMpc$, presented a wide choice of triangular configurations of different shapes and obtained a determination of the bispectrum with an overall errors of the order of $3$-$4\%$. Of particular interest in this range of scales are the nonlinear corrections induced by gravitational instability {\em due} to non-Gaussian initial conditions as they generate an additional non-Gaussian signal on top of the primordial component. For a nonlinear parameter $\fNL=100$, we found that the amplitude of these corrections range from $3$-$4\%$ for generic triangle configurations up to $20$-$30\%$ for ``squeezed'' configurations where we expect most of the signal for local non-Gaussianity. We quantified these corrections with the aid of the ratio and the difference between the non-Gaussian and the Gaussian bispectrum. Our set of eight different realizations of those models ensure that our results are robust to sampling variance. We considered simulations snapshots at redshift $z=0$, $1$ and $2$. Overall, we found that the magnitude of the correction induced by non-Gaussian effects is similar regardless the scale and the redshift. This is due to a compensation between the primordial component that decreases with time on the one hand, and the contribution from nonlinear structure growth that increases with time on the other hand.

We compared our results with the predictions of Eulerian perturbation theory, both at tree-level and one-loop \citep{Sefusatti2009}. As expected, and similarly to what happens for Gaussian initial conditions, the tree-level approximation fails at relatively large scales, $k\sim0.05$ - $0.1\kMpc$, even at high redshift. One-loop corrections extend significantly the predictive power of perturbation theory down to mildly non-linear scales $k\sim 0.3\kMpc$ at redshift $z\gtrsim 1$, similarly to the case of the power spectrum analyzed by\citet{JeongKomatsu2006}. They describe, in fact, the matter bispectrum measured in simulations at the few percent level, with an even better agreement with respect to the ``relative'' effect of primordial non-Gaussianity on the Gaussian bispectrum. Furthermore, they also show a good qualitative agreement with simulations at redshift zero.

\section*{Acknowledgments}

We thank Roman Scoccimarro and Francis Bernardeau for useful comments. E.~S.\ acknowledges support by the French Agence National de la Recherche (ANR) under grant BLAN07-1-212615 and by the European Commission under the Marie Curie Inter European Fellowship and he is grateful to the Institute for Theoretical Physics of the University of Z\"urich for hospitality during the completion of this project. M.~C.\ would like to thank the Institut de Physique Th\'eorique of CEA-Saclay for hospitality and acknowledges support from the Spanish Ministerio de Ciencia y Tecnologia (MEC) through the Juan de la Cierva program. V.~D.\ would like to thank the Institut d'Astrophysique de Paris for hospitality during the final stages of this work and the Swiss National Foundation (under contract No. 200021-116696/1) for support.

\label{lastpage}

\end{document}